\documentclass[twocolumn,pra]{revtex4}
\usepackage{graphicx}

\begin{document}
\topmargin-1cm

\title {Time-dependent $V$-representability on lattice systems}

\author {Yonghui Li}
\author {C. A. Ullrich}
\affiliation {Department of Physics and Astronomy, University of Missouri, Columbia, Missouri 65211, USA}

\date{\today}

\begin{abstract}
We study the mapping between time-dependent densities and potentials for noninteracting electronic systems on lattices. As discovered recently
by Baer [J. Chem. Phys. {\bf 128}, 044103 (2008)], there exist well-behaved time-dependent density functions on lattices which cannot be
associated with any real time-dependent potential. This breakdown of time-dependent $V$-representability can be tracked down
to problems with the continuity equation which arise from discretization of the kinetic-energy operator. Examples are
given for lattices with two points and with $N$ points, and implications for practical numerical applications of time-dependent
density-functional theory are discussed. In the continuum limit, time-dependent noninteracting $V$-representability is restored.
\end{abstract}

\maketitle

\section{Introduction} \label{sec:intro}

Time-dependent density-functional theory (TDDFT) \cite{TDDFTbook} is a widely used methodology
for the description of electron dynamics in atoms, molecules and solids. The birth of TDDFT dates
back to the seminal work by Runge and Gross (RG) \cite{Runge1984} who proved a one-to-one correspondence
between time-dependent densities, $n({\bf r},t)$, and time-dependent potentials $V({\bf r},t)$. The
RG theorem \cite{Runge1984} can be viewed as the time-dependent analog of the Hohenberg-Kohn
theorem \cite{HK1964} in static DFT \cite{Kohn1999}.

While TDDFT has found widespread practical use for describing such diverse phenomena as molecular excitations, nanoscale electron
transport, or strong-field processes, there is a continuing need to study the fundamental underpinnings
of the theory. Some of the basic issues in TDDFT that have recently received much attention are
nonadiabaticity and memory effects \cite{Wijewardane,UllrichTokatly}, and causality and the time-dependent variation
principle \cite{Vignale2008}.

This paper has been motivated by an unexpected and disquieting discovery made recently by Baer \cite{Baer2008}:
on lattice systems there exist seemingly well-behaved time-dependent density functions which are not $V$-re\-pre\-sen\-ta\-ble (VR),
i.e., which cannot be associated with any time-dependent potential.
Attempts to invert the time-dependent Schr\"odinger equation to
explicitly construct the potential from such densities encounter severe numerical instabilities. It was suggested
in Ref. \cite{Baer2008} that such
instabilities might be potentially disastrous for practical applications of TDDFT, since they could
cause the exchange-correlation (XC) potential $V_{\rm xc}$ to become an extremely sensitive functional of the density.

The purpose of this paper is to shed light on two important questions which are raised by Baer's study:
(1) what is the fundamental reason for the
absence of time-dependent $V$-representability on lattices, and (2) how does time-dependent $V$-representability emerge in the continuum limit?
Both questions have a profound impact on our fundamental understanding of TDDFT as well as on practical applications.
We will give some general answers, as well as some specific illustrations for the mapping between time-dependent potentials
and noninteracting densities on one-dimensional lattices.

\section{DFT on lattice spaces} \label{sec:DFT}

The basic existence theorem of ground-state DFT, the (first) HK theorem \cite{HK1964}, states that there is a 1:1 correspondence
between external potentials and ground-state densities. In other words, it cannot happen that two potentials that
differ by more than a constant produce the same ground-state density of an $N$-particle system. The proof of this
theorem is based on the Ritz variational principle:
\begin{equation}
\langle \Psi_{\rm gs} | \hat{H} | \Psi_{\rm gs} \rangle < \langle \Psi | \hat{H} | \Psi \rangle ,
\end{equation}
where $\hat{H}$ is the Hamiltonian of the interacting many-particle system, $\Psi_{\rm gs}$ is the associated ground-state wave function, and
$\Psi$ is any other many-particle wave function. Notice that this fundamental variational
principle does not require that the system lives in a continuous coordinate space; it remains valid for systems that are defined on
a discrete space, such as on a real-space lattice or using a finite set of basis functions.

The basic DFT framework assumes interacting and noninteracting $V$-representability \cite{dreizlergross}. In other words, it is assumed
that any density function which is mathematically reasonable (i.e., does not diverge, and integrates to $N$) is a
ground-state density belonging to some external potential. This property is of obvious importance for a mathematically meaningful implementation
of variations with respect to the density. Unfortunately, attempts to prove $V$-representability for all densities in
the continuum case (e.g. in Ref. \cite{Chen}) have so far not been fully successful.

In ground-state DFT,
lattice systems have been much used to address the $V$-representability problem. In 1983,
Kohn \cite{Kohn1983} showed that in the vector space of density functions that are defined on discrete lattices,
a density in the neighborhood of a VR density is also VR. In 1985, Chayes, Chayes and Ruskai
\cite{Chayes1985} proved that any mathematically well-behaved density function on finite or infinite lattices
can be represented as the density of a pure ground state or as an ensemble density associated with a degenerate ground state.
In density vector spaces of arbitrary dimension, it is equally likely to encounter pure-state and ensemble-state VR
densities \cite{Ullrich2002}. Since the lattice can be arbitrarily dense, this effectively solves the $V$-representability problem in DFT.
In a related approach, $V$-representability was recently proved by Lammert for coarse-grained systems \cite{Lammert2006}.

We mention that lattice systems have also been very helpful in studying the issue of nonuniqueness in spin-DFT, where
it was discovered \cite{Ullrich2005} that there are several different types of situations where the mapping between
densities and spin magnetizations, $\{n({\bf r}), {\bf m}({\bf r})\}$, and potentials and magnetic fields,
$\{V({\bf r}),{\bf B}({\bf r})\}$, is not unique.
In the continuum limit, only some special cases of nonuniqueness in spin-DFT survive.

\section{$V$-representability in continuous- and discrete-space TDDFT} \label{sec:TDDFT}

The issue of $V$-representability in TDDFT turns out to be quite different from static DFT.
Van Leeuwen (vL) \cite{RvL1999} showed that, assuming that the initial state is noninteracting VR, there is always
a unique time-dependent potential $V({\bf r},t)$ in a noninteracting system that produces a given interacting
density $n({\bf r},t)$ at all times. In other words, the time-dependent $V$-representability problem
in the continuum case can be considered solved by the vL-construction. In view of this, the finding by
Baer \cite{Baer2008} that time-dependent noninteracting $V$-representability breaks down on lattices is particularly unsettling
and calls for an explanation.

Unlike the HK theorem in static DFT, the RG theorem \cite{Runge1984} and the vL construction \cite{RvL1999} are not based
on a minimum principle. Instead, it is demonstrated that potentials which differ from each other by more than a time-dependent
function cause the time evolution of the respective systems to proceed in such a manner that the time-dependent densities are not the
same. The original RG proof requires two steps: first it is shown that if a system evolves from a given initial state under the influence
of two external potentials $V ({\bf r},t)$ and $V'({\bf r},t) \ne V({\bf r},t) + c(t)$ then
the current densities ${\bf j}({\bf r},t)$ and ${\bf j}'({\bf r},t)$
are different. The next step shows  by virtue of the continuity equation
that the associated particle densities $n({\bf r},t)$ and $n'({\bf r},t)$ become different, too.
The continuity equation also plays a crucial role in the vL construction.

Let us therefore take a closer look at the continuity equation, given by
\begin{equation} \label{continuity}
\dot{n}({\bf r},t) = -\nabla \cdot {\bf j}({\bf r},t) \:,
\end{equation}
where the dot denotes a partial derivative with respect to time. Here and in the following, we use Hartree atomic units
with $\hbar = e = m = 1$. For the simplest case, a single particle in one dimension, the continuity equation becomes
\begin{equation} \label{continuity1}
\dot{n}(x,t) = \frac{i}{2} \left[ \psi^*(x,t) \frac{\partial^2 \psi(x,t)}{\partial x^2} -
\psi(x,t) \frac{\partial \psi^*(x,t)}{\partial x^2} \right].
\end{equation}
Let us now treat the time variable as continuous but the space variable $x$ as discretized on a grid with equidistant
grid spacing $a$. The position of the $j$th lattice point is $x_j$, and $n_j$ denotes the number of particles in a bin of width $a$,
centered at $x_j$. Thus, the lattice quantity corresponding to the particle density $n(x,t)$ is $n_j(t)/a$. The advantage
of this definition is that the normalization condition on an $N$-point lattice becomes simply
\begin{equation}\label{latticenorm}
\sum_{j=1}^N n_j = 1 \:.
\end{equation}
For the wave functions, we have $\psi(x,t) \longrightarrow \psi_j(t)/\sqrt{a}$.

Using a three-point finite-difference representation of the Laplacian operator (which is
accurate to within terms of order $a^2$) \cite{Abramowitz}, we have
\begin{equation}
\dot{n}_j = \frac{i}{2a^2} \psi_j^* \left[ \psi_{j+1} - 2\psi_j + \psi_{j-1}\right]  + c.c. \:,
\end{equation}
where the time arguments are suppressed for brevity.
Without loss of generality we take the following form of the wave function:
\begin{equation} \label{ansatz}
\psi_j = \sqrt{n_j} \: e^{i\alpha_j} \:,
\end{equation}
where $\alpha_j$ is a real-valued time-dependent phase. The discretized continuity equation then becomes
\begin{eqnarray}
 \dot{n}_j &=& \frac{1}{a^2} \sqrt{n_j n_{j+1}} \sin(\alpha_j - \alpha_{j+1}) \nonumber \\
&+&\frac{1}{a^2} \sqrt{n_j n_{j-1}} \sin(\alpha_j - \alpha_{j-1}) \:.
\end{eqnarray}
We will now show that $\dot{n}_j$ is not allowed to take on arbitrary values. We have
\begin{equation}
 |\dot{n}_j| \le \frac{1}{a^2} \left[ \sqrt{n_j n_{j+1}} + \sqrt{n_j n_{j-1}} \right] .
\end{equation}
Next, let $n_{j+1} = n_j + \Delta$. Using the normalization condition (\ref{latticenorm}) one obtains
$n_{j-1} \le 1-2n_j - \Delta$ (the equal sign arises for lattices consisting only of three points,
or if the potential is infinity on all but the three points $j-1,j,$ and $j+1$). Thus,
\begin{equation}
 |\dot{n}_j| \le \frac{\sqrt{n_j}}{a^2} \left[ \sqrt{n_j +\Delta} + \sqrt{1-2n_j-\Delta} \right] .
\end{equation}
The right-hand side is maximized for $\Delta=0$ and $n_j = (3 + \sqrt{3})/12$, which leads to
\begin{equation}\label{1Dcondition}
|\dot{n}_j| \le \frac{0.683}{a^2} \:.
\end{equation}
On a one-dimensional, equidistant grid with a three-point finite-difference representation of the kinetic energy operator,
the time derivative of the density is restricted by the local upper bound (\ref{1Dcondition}).
In other words, the density on a given grid point cannot change arbitrarily fast.

The upper bound (\ref{1Dcondition}) can be generalized for a $d$-dimensional grid using a $k$-point formula for the Laplacian
operator \cite{Abramowitz}, resulting in  $|\dot{n}_j| \le A_k^d/a^2$, where $A_k^d$ is a finite numerical constant.
It is easy to see that on an equidistant $d$-dimensional grid, $A_k^d = d A_k^1$.
Going beyond three-point formulas, one finds for instance $A_5^d = (4/3) A_3^d$ for five-point formulas, and $A_k^d \to 2 A_3^d$ for large $k$.

Thus, there are fundamental limits, determined by the grid spacing, the number of dimensions, and the discrete representation of the
kinetic-energy operator, on how fast lattice densities can locally change without violating the continuity equation.
The RG theorem and the vL construction are thus
restricted to time-dependent densities which satisfy these bounds. Otherwise, time-dependent $V$-representability cannot be guaranteed.
We will now illustrate this with some explicit examples.

\section{The 2-point lattice} \label{sec:twopoint}

The case of the 2-point lattice was already treated by Baer \cite{Baer2008}; nevertheless, it is a very instructive and pedagogical model,
and we will discuss it here in some detail.

We start with the single-electron time-dependent  Schr\"o\-dinger equation (TDSE) in real space:
\begin{equation}\label{tdse}
i\frac{\partial}{\partial t}\psi({\bf r},t) = \left[-\frac{\nabla^2}{2} + \tilde{V}({\bf r},t)\right]\psi({\bf r},t) \:.
\end{equation}
On a two-point lattice, the wave function is completely specified by its values on point 1 and on point 2, $\psi_1(t)$ and $\psi_2(t)$. These
two (complex) quantities can be arranged as the components of a vector.
The TDSE can then be discretized and written in the form of a $2\times 2$ matrix equation:
\begin{equation} \label{tdse1}
i \frac{\partial}{\partial t}\left( \begin{array}{c} \psi_1(t) \\ \psi_2(t) \end{array}\right)
= \left( \begin{array}{cc} \frac{1}{a^2} + \tilde{V}_1(t) & -\frac{1}{2a^2} \\[1mm]
 -\frac{1}{2a^2} & \frac{1}{a^2} + \tilde{V}_2(t) \end{array} \right)
\left( \begin{array}{c} \psi_1(t) \\ \psi_2(t) \end{array}\right),
\end{equation}
where we used a standard 3-point finite-difference representation of the kinetic-energy operator assuming
that the two lattice points have a spacing $a$.

The notation can be simplified recognizing that
an overall constant shift of the potential is irrelevant. This means that the potential can be redefined as
$V_j = \frac{1}{a^2}+ \tilde{V}_j$, $j=1,2$, and
\begin{equation} \label{tdse2}
i \frac{\partial}{\partial t}\left( \begin{array}{c} \psi_1(t) \\ \psi_2(t) \end{array}\right)
= \left( \begin{array}{cc} V_1(t) & -\frac{1}{2a^2} \\ -\frac{1}{2a^2} & V_2(t) \end{array} \right)
\left( \begin{array}{c} \psi_1(t) \\ \psi_2(t) \end{array}\right).
\end{equation}
We now write the wave function on points 1 and 2 in the form (\ref{ansatz}).
The densities and phases, $n_{1,2}$ and
$\alpha_{1,2}$, are real functions of time. Inserting this in the TDSE (\ref{tdse2}) gives the following equations
for the real and imaginary part of the potential on point 1:
\begin{eqnarray}
\Re V_1 &=& -\dot{\alpha}_1 + \frac{1}{2a^2} \sqrt{\frac{n_2}{n_1}} \:\cos(\alpha_2 - \alpha_1)  \label{REV}\\
\Im V_1 &=& \frac{\dot{n}_1}{2n_1} + \frac{1}{2a^2} \sqrt{\frac{n_2}{n_1}} \:\sin(\alpha_2 - \alpha_1) \:, \label{IMV}
\end{eqnarray}
and similar on point 2.
If we require the potential to be purely real, then the imaginary part, given by equation (\ref{IMV}), vanishes.
We can then solve for the phase difference:
\begin{equation} \label{9}
\alpha_1 - \alpha_2 = \sin^{-1}\left(\frac{a^2 \dot{n}_1}{\sqrt{n_1 n_2}}\right).
\end{equation}
Due to norm conservation (\ref{latticenorm}), we have only one independent density variable, since $n_1 = 1 - n_2$.
Likewise, there is only one independent potential variable, since an arbitrary overall constant shift only affects the phases \cite{Runge1984}.
Thus, only the potential {\em difference} matters, and we let $\Delta V \equiv \Re V_2 - \Re V_1$. From Eq. (\ref{REV}) we get
\begin{equation} \label{DeltaV}
\Delta V = -(\dot{\alpha}_2 - \dot{\alpha}_1) + \frac{1}{2a^2}
\left[ \sqrt{\frac{n_1}{n_2}} - \sqrt{\frac{n_2}{n_1}}\right] \cos(\alpha_1 - \alpha_2) \:,
\end{equation}
and the phase differences can be eliminated using Eq. (\ref{9}). This yields the potential difference as a function of the densities
at point 1 and 2:
\begin{eqnarray}\label{potentialdiff}
\Delta V &=&\frac{a^2}{\sqrt{n_1n_2 - a^4\dot{n}_1^2}} \left[\ddot{n}_1 + \frac{\dot{n}_1^2( n_1 - n_2)}{2n_1n_2}  \right]
\nonumber\\
&+&
\frac{n_1-n_2}{2a^2n_1n_2}\sqrt{n_1n_2 - a^4\dot{n}_1^2} \:.
\end{eqnarray}
This equation for $\Delta V$ is well-behaved and has a unique solution as long as $n_1n_2 - a^4\dot{n}_1^2 >0$, or
\begin{equation} \label{constraint}
|\dot{n}_{1,2}|< \frac{\sqrt{n_1n_2}}{a^2}
\end{equation}
($n_1$ and $n_2$ themselves are of course always $>0$; furthermore, $\dot{n}_1=-\dot{n}_2$). This yields the following upper bound:
\begin{equation} \label{upper2}
|\dot{n}_{1,2}|_{\rm max} = \frac{0.354}{a^2} \:.
\end{equation}
Time-dependent single-particle lattice densities $n_1$,$n_2$ which violate conditions (\ref{constraint}) or (\ref{upper2})
at any time $t$ do not produce a real value of $\Delta V$ at that time and are therefore not time-dependent VR.
This upper bound for the two-point lattice is clearly more restrictive than condition (\ref{1Dcondition}) for a general one-dimensional lattice.

As an illustration, let us consider a one-dimensional time-dependent density function of the form
\begin{eqnarray} \label{nxt}
n(x,t) &=& A_1^2 \cos^2(\pi x)
+A_2^2 \sin^2(2\pi x) \nonumber\\
&+&
2 A_1 A_2 \cos(\pi x) \sin(2\pi x)  \cos(\omega t) \:,
\end{eqnarray}
$-1/2 \le x \le 1/2$,
arising from a superposition of the first and second single-particle eigenstate of a particle in a one-dimensional box of length one.
We discretize $n(x,t)$ on a 2-point lattice such that $n_{1,2}=n(x_{1,2},t)$, where $x_1 = -1/6$ and $x_2=1/6$, and the lattice
spacing is $a=1/3$. The lattice normalization condition (\ref{latticenorm}) requires choosing $A_1$ and $A_2$ such that $n_1+n_2=1$.

\begin{figure}
\unitlength1cm
\begin{picture}(5.0,7)
\put(-8.5,-13){\makebox(5.0,6.5){\includegraphics{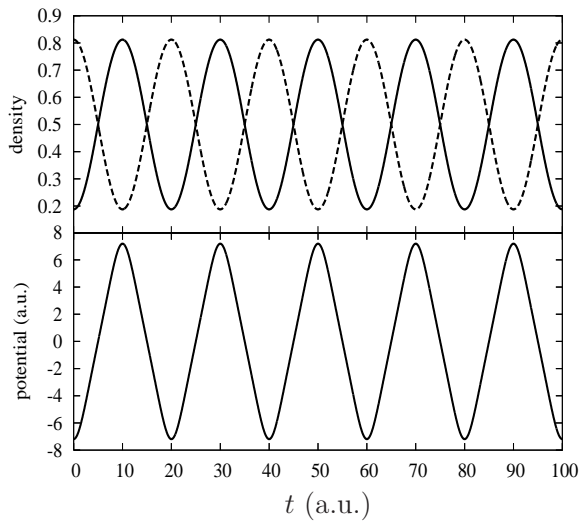}}}
\end{picture}
\caption{\label{figure1}
Top: two-point lattice densities $n_1$ (full line) and $n_2$ (dashed line) resulting from $n(x,t)$ of Eq. (\ref{nxt}), with $\omega=0.1\pi$.
Bottom: potential difference $\Delta V=V_2-V_1$ [Eq. (\ref{potentialdiff})].
}
\end{figure}

The top panel of Fig. \ref{figure1} shows $n_1$ and $n_2$ as a function of time for a frequency $\omega=0.1 \pi$ and
$A_1= 0.77$ and $A_2=0.27$, together with
the associated potential difference $\Delta V$, following from Eq. (\ref{potentialdiff}). This is a case where the lattice densities
are VR, and the potential is a well-behaved function of time which drives the particle density periodically from the first to
the second lattice point and back. In fact, in this low-frequency case the adiabatic approximation for the potential,
\begin{equation} \label{adia}
\Delta V_{\rm adia}(t) = \frac{n_1(t)-n_2(t)}{2a^2 \sqrt{n_1(t)n_2(t)}}
\end{equation}
(which assumes that at each moment in time the system is in the ground state associated with the instantaneous potential),
is indistinguishable from the exact $\Delta V(t)$.

\begin{figure}
\unitlength1cm
\begin{picture}(5.0,15)
\put(-8.5,-7.3){\makebox(5.0,6.5){\includegraphics{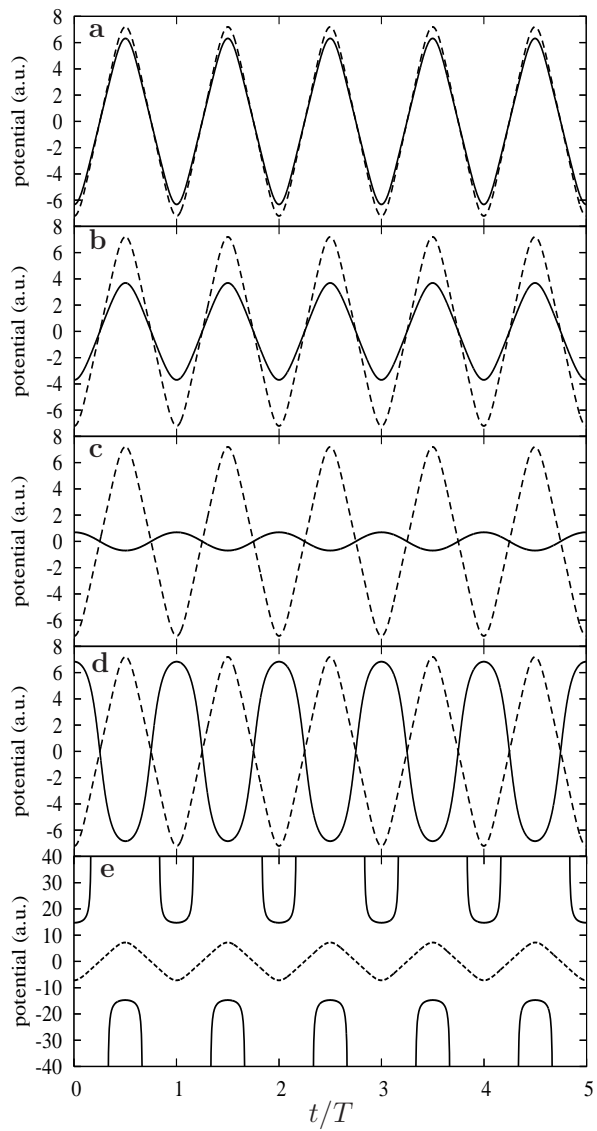}}}
\end{picture}
\caption{\label{figure2}
Full lines: potential difference $\Delta V=V_2-V_1$ for $\omega = \pi, 2\pi, 3\pi, 4\pi$, and $5\pi$ (a-e), with $T=2\pi/\omega$.
Dashed lines: adiabatic potential [Eq. (\ref{adia})]. Panel (e) shows regions where no potential exists that
produces the given lattice densities $n_1(t)$ and $n_2(t)$, which means that they are non-VR.
}
\end{figure}

Figure \ref{figure2} shows the potential reproducing $n_1$ and $n_2$ at higher frequencies, from $\omega= \pi$ through $5\pi$.
As $\omega$ grows, the time-dependent potential $\Delta V$ starts to differ more and more from the adiabatic approximation
$\Delta V_{\rm adia}$. In fact, there is a crossover at $\omega=2.865 \pi$ where one finds
$\Delta V(t) \equiv 0$, which corresponds to free charge-density oscillations on the two-point lattice with frequency $\omega=1/a^2$
(the difference of the two eigenvalues of the static Schr\"odinger equation with $V_1=V_2=0$). For $\omega>2.865 \pi$,
$\Delta V$ and $\Delta V_{\rm adia}$ are out of phase, which may be viewed as an indication of entering the high-frequency regime.

For $\omega > 4.587\pi$ one finds that $\Delta V$ diverges periodically, and there are regions in time where there is no solution at all,
as seen for $\omega = 5\pi$ in panel (e) of Figure \ref{figure2}.
This is a striking example where the time-dependent lattice density is non-VR.

In the regions in which $\Delta V$ does not exist, condition (\ref{constraint}) is violated, which implies that the densities change
too rapidly. In other words, there is a limit on how much density can move how fast between the two points.
Violation of this limit means that no real-valued time-dependent lattice potential can be found which is capable of driving the system
strongly and fast enough to produce the given time-dependent density.

To conclude this section, we mention that some of the issues addressed here have also
been recently discussed by Verdozzi in the context of Hubbard dimers \cite{Verdozzi}.

\section{The $N$-point lattice}
For a linear $N$-point lattice with equidistant grid spacing $a$ and a 3-point finite-difference formula, the discretized
TDSE reads
\begin{equation} \label{tdseN}
i\left( \begin{array}{c} \dot{\psi}_1 \\ \dot{\psi}_2  \\ \dot{\psi}_3 \\ \vdots \\ \dot{\psi}_N \end{array}\right)
= \left( \begin{array}{ccccc}
 V_1 & -\frac{1}{2a^2} & 0 &  \ldots & 0 \\
 -\frac{1}{2a^2} & V_2 & -\frac{1}{2a^2} & \ldots & 0 \\
0 & -\frac{1}{2a^2} & V_3 & \ldots & 0 \\
\vdots & \vdots & \vdots& \ddots & \vdots \\
0 & 0 &  0 & \ldots & V_N
 \end{array} \right)
\left( \begin{array}{c} \psi_1 \\ \psi_2 \\ \psi_3 \\ \vdots\\ \psi_N \end{array}\right),
\end{equation}
where the time arguments of the $\psi_i$ and $V_i$ have been omitted and the
constant $1/a^2$ coming from the finite-difference kinetic-energy operator has been absorbed in the
potentials $V_i(t)$, as in Eq. (\ref{tdse2}).

It is straightforward to generalize the analysis of the previous section to the $N$-point lattice. We again use ansatz (\ref{ansatz})
and eliminate the phases using the requirement that the potential be real at each point.
Defining
\begin{equation} \label{Sk}
S_k = \sum_{j=1}^{k} \dot{n}_j \:,
\end{equation}
we obtain after some algebra the following expression for the point-to-point potential differences:
\begin{widetext}
\begin{eqnarray} \label{Npoint}
V_{k+1}-V_k &=& \frac{a^2}{\sqrt{n_k n_{k+1} - a^4 S_k^2}}
\left[ \dot{S}_k - \frac{(n_k\dot{n}_{k+1} + n_{k+1}\dot{n}_k) S_k}{2n_{k} n_{k+1}}
\right]
+ \frac{n_k - n_{k+1}}{2a^2n_k n_{k+1}}\sqrt{n_k n_{k+1} - a^4 S_k^2}
\nonumber\\
&+&
\frac{1}{2a^2n_{k+1}}\sqrt{n_{k+1} n_{k+2} - a^4 S_{k+1}^2}
- \frac{1}{2a^2n_{k}}\sqrt{n_{k-1} n_{k} - a^4 S_{k-1}^2} \;, \qquad k=1,\ldots, N-1.
\end{eqnarray}
\end{widetext}
For notational convenience,
Eq. (\ref{Npoint}) makes reference to the zeroth and the $(N+1)$st lattice point. These points should be
regarded as ``virtual points'', consistent with the boundary condition $n_0 = n_{N+1} = 0$.
With this and the normalization condition (\ref{latticenorm}) it is easy to see that Eq. (\ref{Npoint}) reduces
to Eq. (\ref{potentialdiff}) for $N=2$.

Eq. (\ref{Npoint}) yields real solutions for the point-to-point potential differences $V_{k+1}-V_k$
as long the terms under the square
roots remain positive. This immediately leads to the following constraint on the time-dependent lattice density, which
is a generalization of Eq. (\ref{constraint}):
\begin{equation}\label{critical}
|S_k | < \frac{\sqrt{n_k n_{k+1}}}{a^2}\:, \qquad k=1,\ldots, N-1.
\end{equation}
This condition represents a much tighter and more specific criterion for the $V$-representability
of a given time-dependent lattice density than the upper bound (\ref{1Dcondition}).

Let us compare the two criteria for the same oscillating density function (\ref{nxt}), with $A_1=0.77$ and $A_2=0.27$,
that was studied in section \ref{sec:twopoint}. We consider $N$-point lattices with grid spacing $a=1/(N+1)$ and lattice points at
$x_j = ja-1/2$, $j=1,\ldots,N$, and we normalize the discretized density using (\ref{latticenorm}).

\begin{figure}
\unitlength1cm
\begin{picture}(5.0,5.75)
\put(-9.7,-15.9){\makebox(5.0,6){\includegraphics{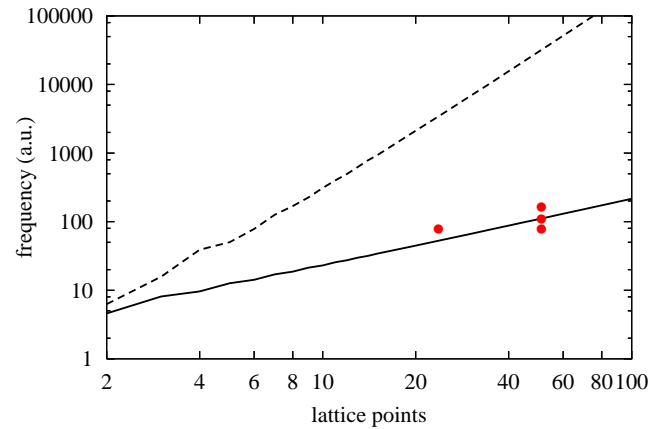}}}
\end{picture}
\caption{\label{figure3}
Full line: critical frequency at which $n(x,t)$ from Eq. (\ref{nxt}) becomes non-VR on an $N$-point lattice. Dashed line: frequency at
which the density would violate the upper bound (\ref{1Dcondition}). The dots indicate the cases studied in
Figure \ref{figure4}.
}
\end{figure}

The full line in figure \ref{figure3} shows the critical frequency at which the time-dependent density becomes non-VR on at least one
point on an $N$-point lattice, according to Eq. (\ref{critical}). By comparison, the dashed line shows the frequency at which the same
time-dependent density would start to violate
the upper bound (\ref{1Dcondition}). One finds that the former increases linearly with the number of lattice points $N$, whereas the latter
grows as $N^3$. This drastically different behavior is not surprising: the absolute upper bound (\ref{1Dcondition}) corresponds to the extreme limit
where the density is completely concentrated on three lattice points only. For $N<4$, this gives a quite good agreement
as seen in figure \ref{figure3}, but the more points one adds to the lattice, the less this extreme scenario applies to the
density that we actually consider here.

\begin{figure}
\unitlength1cm
\begin{picture}(5.0,12.2)
\put(-9.1,-12.1){\makebox(5.0,12.2){\includegraphics{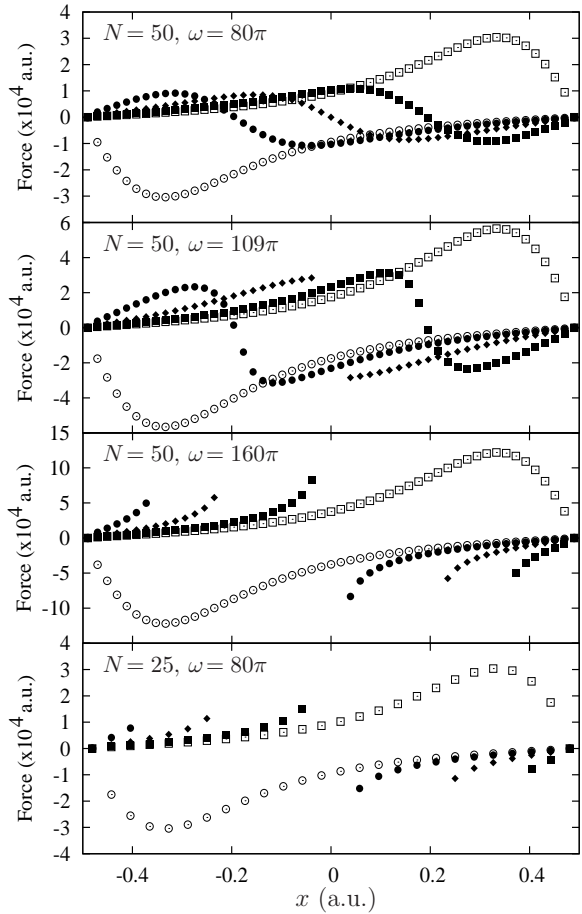}}}
\end{picture}
\caption{\label{figure4}
Snapshots of the time-dependent force (\ref{force}) generating the density (\ref{nxt})
for $t=0$ (open circles), $T/8$ (full circles), $T/4$ (diamonds),
$3T/8$ (full squares), and $T/2$ (empty squares). Top panel: force exists everywhere at $N=50$ and $\omega=80\pi$,
$n(x,t)$ is lattice-VR.
Other panels: $V$-representability breaks down for increasing $\omega$ at fixed $N$, or
decreasing $N$ at fixed $\omega$.}
\end{figure}

Figure \ref{figure4} gives further evidence of the breakdown of $V$-representability. We plot the time-dependent
force associated with the potential differences (\ref{Npoint}),
\begin{equation} \label{force}
F_{k+\frac{1}{2}} = -( V_{k+1}-V_k)/a \,,
\end{equation}
which represents a discretized version of the continuum expression $F(x) = - \partial V(x)/\partial x$.
At $N=50$ and $\omega=80\pi$ the density (\ref{nxt}) is VR, as can be seen from Figure \ref{figure3} (the dot
below the full line). The forces $F_{k+\frac{1}{2}}$ are therefore well behaved, i.e., they exist and are finite everywhere on the
lattice at all times. If we keep the number of lattice points fixed but increase the
frequency, we cross the line of critical frequencies at $\omega=109 \pi$:
$V$-representability breaks down and there are regions on the lattice and times during the cycle
where no force exists. A similar breakdown happens for fixed frequency if the number of lattice points is reduced.

\section{Discussion}

From the above analysis of one-electron lattice densities, we can draw a number of conclusions about noninteracting
$V$-representability in TDDFT.

{\em Continuum limit.} -- Condition (\ref{critical}), as well as the upper limit (\ref{1Dcondition}), both become more and more easily
satisfied as the grid spacing $a$ decreases. This is also evident from Figure \ref{figure3}. Thus, in the continuum limit $a\to 0$,
which corresponds to a lattice whose distribution of grid points is infinitely dense, all time-dependent single-particle densities become VR.

There are various ways to see how time-dependent $V$-representability emerges in the continuum limit. First of all, as discussed
in section \ref{sec:TDDFT}, the continuity equation is in general not valid on lattices, which breaks crucial links in the chains of
arguments that constitute the RG proof and the vL construction. In the $a\to 0 $ limit, the continuity equation is restored, and thus the fundamental
proofs of TDDFT which establish the one-on-one mapping between densities and potentials.

Another way of looking at things is to consider the energy eigenvalue spectrum associated with a particular lattice.
Each excitation energy can be associated with a resonant electronic eigenmode, i.e., a free (undriven) oscillation of arbitrary amplitude.
If the time dependence of a given lattice density is close to that of one of the lattice eigenmodes, or a superposition thereof,
then it is possible to find an external time-dependent driving potential which gives rise to this density.
On the other hand, if the given time-dependent density changes much more rapidly than any available lattice mode, then
no external potential exists which is capable of inducing these rapid changes. This was clearly demonstrated in section \ref{sec:twopoint}
for the 2-point lattice.

Ultimately, it is the time-energy uncertainty relation $\Delta t \Delta E \agt \hbar$ which governs how fast a quantum state can change
in time \cite{Pfeifer}. Here $\Delta E$ indicates a spectral measure of the initial Hamiltonian of the system, such as its ``energy spread''.
Consider a 1-dimensional $N$-point lattice with grid spacing $a$, and uniform (constant) potential. There are $N$ eigenvalues
$\varepsilon_n$, $n=1,\ldots,N$, of the static Schr\"odinger equation, and $N-1$ eigenmodes with frequency
$\omega_n = \varepsilon_n - \varepsilon_1$. The dynamical range or energy spread of this lattice is given by the difference
of the highest and the lowest eigenvalue, which scales as $1/a^2$. According to the uncertainty relation, the range of possible
response times of the system decreases at the same rate. Thus, the smaller the grid spacing, the more rapidly the system
can change in time, and the wider the range of time-dependent $V$-representability.

{\em Complex potentials.} -- It may be of interest to observe that non-VR lattice densities can still be associated with external potentials.
However, these potentials have to be complex, and are therefore non-Hermitian. Complex potentials are a well-known tool to
generate absorbing boundary conditions for quantum dynamics simulations on finite grids \cite{Vibok1992}.
However, complex time-dependent potentials can also be constructed to be norm-conserving, which means that the resulting time evolution
is unitary. This is explicitly demonstrated in the Appendix.

The physical mechanism by which a complex potential reproduces a non-VR density is not by
``driving'' the density from one lattice point to another, as a real potential would try to do. Rather, complex potentials introduce
local particle sources and sinks. A time-dependent density that moves rapidly between two lattice points can thus be viewed as being
``destroyed'' on the first point and ``re-created'' on the second point. There is in principle no limit as to how fast such processes can take place.
This is also reflected in the fact that the continuity equations (\ref{continuity}) and (\ref{continuity1}) are modified for complex
potentials to include extra source terms.

The mapping between real densities (a single variable) and complex potentials (two variables: real and imaginary part) obviously cannot be unique,
as shown in the 2-point lattice example in the Appendix. It may be possible to restore uniqueness of the mapping by including the phase
information of the wave function, or equivalently by using the current as an additional variable.

{\em Many-electron systems.} -- In this paper, we have only studied the case of a single electron in detail. However, the essential arguments for and
against noninteracting $V$-representability on lattices, see section \ref{sec:TDDFT},
can be carried over to the case of many electrons.
In the time-dependent Kohn-Sham (TDKS) scheme, the continuity equation holds for each individual Kohn-Sham orbital, and it is immediately
seen that this can be violated on a finite lattice. The upper bound (\ref{1Dcondition}) can be generalized to the case of
several electrons in a straightforward manner. It is thus clear that time-dependent $V$-representability for many-electron system is only
guaranteed in continuous space, and not on lattices.

It is possible to invert
the TDKS equation for two electrons in a doubly occupied orbital, which was recently done for Hooke's atom
\cite{Hooke,Hessler2002,Wijn2007} and
two-dimensional quantum strips \cite{Ullrich2006}. However, for more than two electrons the construction of the potential in terms
of the density can no longer be carried out analytically [such as in Eq. (\ref{Npoint})], but only numerically.

{\em Ensemble-VR densities.} -- In static DFT, examples of non-VR densities were first constructed by Levy \cite{Levy1982}. The conundrum was
soon resolved by recognizing that such densities can come from ensembles of degenerate states. This finally led to the modern view of
$V$-representability in DFT in terms of pure- and ensemble-state VR densities, see the discussion in section \ref{sec:DFT}.

The role played by degeneracies in TDDFT has so far not been studied in much detail; the situation is much more complex than in
static DFT. However, it is highly unlikely that the absence of $V$-representability on lattices can be cured by the same trick that
worked in static DFT, that is, by representing non-VR densities via ensembles of degenerate states. The simple reason is that, unlike in
static DFT, non-VR time-dependent densities occur already for one-electron systems. Furthermore, the physical origins of
non-$V$-representability in TDDFT are fundamentally tied to the dynamics (such as the continuity equation and how fast quantum states can
change -- see the above discussion).

{\em Practical consequences.} -- In Ref. \cite{Baer2008} it was argued that $V_{\rm xc}$ is likely to be an extremely
sensitive functional of the time-dependent density, especially in strongly time-dependent problems,
due to the possible occurrence of instabilities and breakdown of the density-on-potential mapping on lattices.
Indeed, $V_{\rm xc}$ is defined as a functional of the time-dependent density on the domain of densities that are
noninteracting VR, and is therefore undefined for non-VR densities. However, does this have any practical consequences for TDDFT?

$V_{\rm xc}$ needs to ``know'' whether a time-dependent lattice density is noninteracting VR or not.
Continuum-space XC functionals are not required to have this property, since all time-dependent continuum densities are noninteracting
VR by the vL construction \cite{RvL1999} (as long as the initial state is VR).
Therefore, the exact or any approximate $V_{\rm xc}$  from {\em continuum} TDDFT, if used in a {\em lattice} TDKS calculation,
remains well defined (real and finite) even if evaluated with non-VR lattice densities.
To deal properly and consistently with non-VR lattice densities, $V_{\rm xc}$ must be
lattice dependent, i.e., it needs to be constructed for many-body systems on specific lattices.

Let us now assume that we have the exact $V_{\rm xc}$ for a specific lattice system, and we want to carry out
a self-consistent TDKS calculation using an iterative procedure such as discussed in Ref. \cite{Wijewardane2008}.
The first iteration step involves evaluating $V_{\rm xc}$ with a time-dependent trial density. If this trial
density turns out to be non-lattice-VR, then the TDKS calculation breaks down. If, on the other hand, the trial density
is lattice-VR, then each subsequent iteration produces only lattice-VR densities, and the TDKS calculation will be well-behaved.

Problems could still arise if $V_{\rm xc}$ were extremely sensitive {\em within} the VR region: for example, if it
fluctuated wildly for VR lattice densities that are close to the non-VR region.
However, the examples which we have studied here indicate that this is unlikely to be the case.
Figure \ref{figure4} shows that the force, and hence the potential itself, remains finite and well-behaved as long as
the density is VR, and otherwise simply ceases to exist.

Finally, it is an obvious statement that in the usual applications of TDDFT the density follows from the potential,
not the other way round, and therefore remains VR by definition. Furthermore, we have seen that non-$V$-representability becomes
increasingly rare in the space of time-dependent lattice densities if the grid spacing is made smaller and smaller by adding
more and more lattice points. We may therefore conclude that the fact that the density-on-potential mapping on lattices can
break down has no practical consequences for TDDFT.

\acknowledgments

This work was supported by NSF Grant No. DMR-0553485 and by Research Corporation. We thank Giovanni Vignale for helpful comments.

\appendix*

\section{Complex potentials}

Let us first revisit the derivation of section \ref{sec:twopoint} for the two-point lattice, but now admitting a time-dependent potential
with nonvanishing imaginary part. We impose norm conservation:
\begin{equation}
0 = \dot{n}_1 + \dot{n}_2
=(\psi_1^* \dot{\psi}_1 + \psi_1 \dot{\psi}_1^*) + (\psi_2^* \dot{\psi}_2 + \psi_2 \dot{\psi}_2^*) \:.
\end{equation}
Making use of the TDSE for $\psi_1$ and $\psi_2$, this leads to
the following condition for the imaginary parts of the potentials on points 1 and 2:
\begin{equation} \label{ImV1V2}
n_1 \Im V_1 + n_2 \Im V_2 = 0 \:.
\end{equation}
From Eq. (\ref{IMV}) we obtain
\begin{equation}
\alpha_1 - \alpha_2 = \sin^{-1} \left(\frac{a^2 \dot{n}_1}{\sqrt{n_1 n_2}} -2a^2 \sqrt{\frac{n_1}{n_2}} \: \Im V_1 \right).
\end{equation}
Taking the time derivative of this and plugging it into equation (\ref{DeltaV}) gives, after some algebra,
\begin{eqnarray}\label{ReV1V2}
\lefteqn{
\Re V_2 -\Re V_1 = \frac{a^2}{\sqrt{ n_1 n_2 - a^4(\dot{n}_1 - 2 n_1 \Im V_1)^2}}}
\nonumber \\
&\times& \left\{\ddot{n}_1 + \frac{\dot{n}_1^2(n_1 - n_2 )}{2n_1n_2}
-\frac{n_2 \dot{n}_1}{n_1^2}\:\Im V_1 - 2 n_1  \Im \dot{V_1} \right\} \nonumber\\
&+&
\frac{n_1 - n_2}{2a^2n_1 n_2}
{\sqrt{ n_1 n_2 - a^4(\dot{n}_1 - 2n_1 \Im V_1)^2}} \:,
\end{eqnarray}
which is a generalization of equation (\ref{potentialdiff}) for nonvanishing imaginary part of the potential.
We saw in section \ref{sec:twopoint} that the density is non-VR if the argument $n_1n_2 - a^4 \dot{n}_1^2$ under the square root in
Eq. (\ref{potentialdiff}) becomes negative. If we admit complex potentials, we have the freedom to
choose $\Im V_1$ in such a way that
\begin{equation} \label{cond}
n_1 n_2 - a^4(\dot{n}_1 - 2 n_1 \Im V_1)^2>0 \:.
\end{equation}
This determines $\Re V_2 - \Re V_1$ and $\Im V_2$ via equations (\ref{ImV1V2}) and (\ref{ReV1V2}).
Clearly, $\Im V_1$ is not unique, but each choice for $\Im V_1$ which satisfies condition (\ref{cond}) will
reproduce the given time-dependent density via the lattice-TDSE (\ref{tdse2}).

The arguments presented here for the 2-point lattice can be extended for the $N$-point lattice in a straightforward
manner. Generalizing equation (\ref{ImV1V2}), we find
\begin{equation} \label{NC}
\sum_{j=1}^N n_j \Im V_j = 0 \:
\end{equation}
as a necessary condition that the complex lattice potential is norm-conserving.
The point-to-point differences of the real parts of the potential, $\Re V_{k+1} - \Re V_k$, can then
be obtained from Eq. (\ref{Npoint})
by replacing $S_k$ in Eq. (\ref{Sk}) by
\begin{equation}
S_k = \sum_{j=1}^k (\dot{n}_j - 2 n_j \Im V_j) \:.
\end{equation}
We can therefore always satisfy the constraint (\ref{critical}) through appropriate choices of the imaginary part of
the lattice potential, consistent with the norm-conservation condition (\ref{NC}). As for the 2-point lattice,
these choices are not unique.


\begin{thebibliography}{99}

\bibitem{TDDFTbook}
{\em Time-dependent density functional theory}, edited by M. A. L. Marques, C. A. Ullrich, F. Nogueira,
A. Rubio, K. Burke, and E. K. U. Gross, Lecture Notes in Physics {\bf 706} (Springer,  Berlin, 2006).

\bibitem{Runge1984}
E. Runge and E. K. U. Gross, Phys. Rev. Lett. \textbf{52}, 997 (1984).

\bibitem{HK1964}
P. Hohenberg and W. Kohn, Phys. Rev. {\bf 136}, B864 (1964).

\bibitem{Kohn1999}
W. Kohn, Rev. Mod. Phys. {\bf 71}, 1253 (1999).

\bibitem{Wijewardane}
H. O. Wijewardane and C. A. Ullrich, Phys. Rev. Lett. {\bf 95}, 086401 (2005) and Phys. Rev. Lett. {\bf 100}, 056404 (2008).

\bibitem{UllrichTokatly}
C. A. Ullrich and I. V. Tokatly, Phys. Rev. B {\bf 73}, 235102 (2006).

\bibitem{Vignale2008}
G. Vignale, arXiv:0803.2727v1 (2008).

\bibitem{Baer2008}
R. Baer, J. Chem. Phys. {\bf 128}, 044103 (2008).

\bibitem{dreizlergross}
R. M. Dreizler and E. K. U. Gross, {\em Density-Functional Theory: An Approach to the Quantum Many-Body Problem}
(Springer, Berlin, 1990).

\bibitem{Chen}
J. Chen and M. J. Stott, Phys. Rev. A {\bf 47}, 153 (1993) and Phys. Rev. A {\bf 44}, 2816 (1991).

\bibitem{Kohn1983}
W. Kohn, Phys. Rev. Lett. {\bf 51}, 1596 (1983).

\bibitem{Chayes1985}
J. T. Chayes, L. Chayes, and M. B. Ruskai, J. Stat. Phys. {\bf 38}, 497 (1985).

\bibitem{Ullrich2002}
C. A. Ullrich and W. Kohn, Phys. Rev. Lett. {\bf 89}, 156401 (2002).

\bibitem{Lammert2006}
P. E. Lammert, J. Chem. Phys. {\bf 125}, 074114 (2006).

\bibitem{Ullrich2005}
C. A. Ullrich, Phys. Rev. B {\bf 72}, 073102 (2005).

\bibitem{RvL1999}
R. van Leeuwen, Phys. Rev. Lett. {\bf 82}, 3863 (1999).

\bibitem{Abramowitz}
M. Abramowitz and I. A. Stegun, {\em Handbook of Mathematical Functions} (Dover, New York, 1965).

\bibitem{Verdozzi}
C. Verdozzi, arXiv:0707.2317v1 (2007) and private communication.

\bibitem{Pfeifer}
P. Pfeifer, Phys. Rev. Lett. {\bf 70}, 3365 (1993);
P. Pfeifer and J. Fr\"ohlich, Rev. Mod. Phys. {\bf 67}, 759 (1995).

\bibitem{Vibok1992}
\'A. Vib\'ok and G. G. Balint-Kurti, J. Phys. Chem. {\bf 96}, 8712 (1992).

\bibitem{Hooke}
I. D'Amico and G. Vignale, Phys. Rev. B {\bf 59}, 7876 (2001).

\bibitem{Hessler2002}
P. Hessler, N. T. Maitra, and K. Burke, J. Chem. Phys. {\bf 117}, 72 (2002).

\bibitem{Wijn2007}
A. S. de Wijn, S. K\"ummel, and M. Lein,
J. Comput. Phys. {\bf 226}, 89 (2007).

\bibitem{Ullrich2006}
C. A. Ullrich, J. Chem. Phys. {\bf 125}, 234108 (2006).

\bibitem{Levy1982}
M. Levy, Phys. Rev. A {\bf 26}, 1200 (1982).

\bibitem{Wijewardane2008}
H. O. Wijewardane and C. A. Ullrich, Phys. Rev. Lett. {\bf 100}, 056404 (2008).

\end{thebibliography}
\end{document}